\documentclass{article}
\usepackage{graphicx} 
\usepackage{xcolor}
\usepackage{csquotes}
\usepackage{url}
\usepackage{subfigure}
\usepackage[utf8]{inputenc}
\usepackage[T1]{fontenc}
\usepackage{amsmath}
\usepackage{titlesec}
\usepackage{forest}
\usepackage{amssymb}
\usepackage[margin=3.5cm]{geometry}
\author{Mathilde Leuridan \and James Hawkes \and Tiago Quintino \and Martin Schultz}
\title{Beyond Standard Datacubes: Extracting Features from Irregular and Branching Earth System Data}

\begin{document}
\maketitle
\begin{abstract}
Earth science datasets are growing rapidly in both volume and structural complexity. They increasingly contain richly labelled data with heterogeneous metadata and complex internal constraints that impose dependencies between variables and dimensions.
Datacubes have become a common abstraction for organising such datasets, but traditional dense and orthogonal datacube models struggle to represent irregular, sparse or branching data spaces efficiently. In particular, gaps in coverage, conditional dimensions and heterogeneous variable definitions are poorly captured by existing approaches. In this paper, we introduce a generalised data hypercube representation based on compressed tree structures, which enables an accurate and compact description of complex data spaces. We describe the design of this representation and analyse its ability to capture sparsity and conditional relationships while remaining efficient to traverse. Using a concrete implementation, we study the performance characteristics of compressed tree data hypercubes and demonstrate their effectiveness as fast, cache-like indices over large backend data stores. Building on this representation, we present an integrated feature extraction system that operates directly on tree-based data hypercubes within the Polytope framework. By embedding data access strategies into the data hypercube abstraction itself, the system enables precise, sub-field data extraction and supports flexible, user-driven access patterns. We evaluate the performance of the integrated system and show how it enables new ways of interacting with complex datasets that are difficult to support using traditional access models. This work bridges the gap between expressive data hypercube models and efficient data access methods. In particular, it provides a unified framework that combines tree-based data representations with feature extraction capabilities. The proposed approach therefore offers a foundation for scalable and user-centric access to large heterogeneous Earth science datasets and highlights directions for future development and deployment.
\end{abstract}
\section{Introduction}
Advances in the fields of earth observation, climate modelling and numerical weather prediction have led to unprecedented volumes of geospatial and meteorological data \cite{big_data_weather, big_data_climate, big_data_eo1}. Modern satellites generate a continuous stream of observations across hundreds of spectral channels and instruments, while weather and climate models produce increasingly detailed forecasts with multiple ensemble members, parameterisations and output frequencies. Foundational standardisation efforts, such as the netCDF data model \cite{netcdf} that is implemented with the Climate and Forecast (CF) \cite{cf} and Climate Model Output Rewriter (CMOR) \cite{cmor} metadata conventions, have enabled efficient representation and exchange of datasets structured primarily along the familiar latitude, longitude, vertical level and time dimensions. However, modern datasets increasingly extend beyond this standard model, introducing additional axes such as height levels, instrument modes or forecast lead times, and thereby creating richly structured yet highly complex data spaces.\\
Whilst this abundance of data offers scientific opportunities, it also creates significant practical challenges \cite{big_data_challenges, big_data_frontier}. Indeed, high-dimensional datasets can be difficult to visualise, reason about and manipulate. Analytical operations that are trivial in low-dimensional settings, such as slicing out subsets, combining variables or mapping between coordinate systems, quickly become cumbersome when the data spans irregular grids, contains conditional dependencies or has gaps arising from instrument geometry or quality control. Beyond representation alone, these difficulties directly impact the ability to filter, query and extract meaningful subsets or features from the data, which are essential steps in most scientific workflows. As a consequence, there is a growing need for computational frameworks that can represent such datasets uniformly, while simultaneously enabling efficient and intuitive data access.\\
One key response to this need has been the development of the datacube abstraction. First originating from the business analytics community, the datacube concept provided a way to model multidimensional tabular data and to support operations such as slicing and pivoting \cite{datacube1, datacube2, rolap_qube}. Over the last decade, this idea has been reinterpreted and expanded within the earth sciences community to accommodate raster and observational datasets \cite{eo_datacube1, eo_datacube2, eo_datacube3, eo_datacube4}. Efforts such as RasDaMan \cite{rasdaman1, rasdaman2} and ESA’s Deep Earth System Data Laboratory (DeepESDL) \cite{deepesdl, EarthSystemDataLab_2025} have in particular demonstrated how datacubes can unify heterogeneous datasets and simplify access to complex archives. However, most implementations primarily focus on organising and exposing the structure and metadata of homogeneous multidimensional data, while higher-level data filtering and feature extraction are often treated as downstream, ad-hoc processes.\\
In the open-source scientific ecosystem, the xarray library \cite{xarray} has further popularised the datacube concept by providing a labeled tensor representation for multidimensional arrays. In meteorology, climate research and remote sensing, xarray has become central to data analysis workflows because it facilitates essential operations on complex datasets. The tensor model underpinning xarray carries a few implicit assumptions however. Amongst them, it assumes that data lies on orthogonal, regularly spaced coordinate axes and forms dense arrays without structural gaps. These assumptions are increasingly strained by modern datasets, which may be sparse, irregular or defined only for particular combinations of dimensions, such as when variables depend on specific instruments or ensemble configurations. In such cases, even standard filtering or extraction tasks may require substantial preprocessing or specific logic outside of the datacube abstraction itself.\\
A potential solution to this is to split data into multiple lower-dimensional datacubes. DeepESDL, for example, stores each geophysical parameter as its own 3- or 4-dimensional cube. However, whilst this is a practical representation, this fragmentation hides relationships between parameters and prevents a unified view of the dataset as a whole. More importantly, it complicates the implementation of generic data extraction workflows that must operate consistently across parameters, dimensions and conditional dependencies. Recognising these limitations, xarray recently introduced DataTrees \cite{datatree1, datatree2}, a hierarchical extension designed to express related datacubes in a tree-like structure. DataTrees offer greater flexibility by allowing branching and grouping, but they still impose structural uniformity. Indeed, branches must share similar dimensional assumptions, coordinate ordering and overall data hierarchy, which limits their ability to support generic filtering and feature extraction across highly irregular data spaces.\\
In this paper, we consider a more general representation of high-dimensional data spaces based on a flexible hierarchical tree structure, designed to capture the coordinate structure of complex, heterogeneous and irregular datasets beyond the constraints of existing datacube frameworks. By encoding dimensional dependencies and structural rules directly in the tree, this approach relaxes assumptions of regularity, completeness and uniformity, while supporting efficient operations such as querying, merging, filtering and traversal even for sparse or non-orthogonal data spaces. Crucially, this representation is intended not only to describe the data, but also to serve as a foundation for end-to-end data extraction workflows.\\
We then briefly describe the Qube \cite{ecmwf-qubed} as a concrete realisation of this compressed tree-like datacube representation, and investigate its performance and potential use in practical applications. In particular, we present a prototype integration of the Qube within the Polytope feature extraction framework \cite{polytope, polytope_pasc}, illustrating how feature extraction can be implemented directly on top of generic datacube representations. This enables a complete workflow in which data representation, filtering and feature extraction are tightly integrated, without imposing restrictive assumptions on the structure of the underlying data. Finally, we discuss this integrated framework, highlighting the benefits of a more general approach to datacubes and its potential applicability across a wide range of data extraction and analysis systems.
\section{Motivation and Background}
In recent years, considerable work has focused on formalising the concept of datacubes and multidimensional data spaces to better accommodate the growing complexity of geospatial datasets. In this section, we motivate the need for such abstractions and review current implementations, considering both how multidimensional data are represented and how they are accessed and used in practice. We discuss why high-dimensional scientific datasets require more systematic representations and examine prior frameworks that have shaped current approaches, drawing throughout on examples from meteorological and climate data. In particular, we highlight how characteristics such as multi-instrument observations, ensemble forecasts, heterogeneous grids and conditional data availability stress traditional datacube assumptions. We also review the data access strategies used in existing frameworks, including querying, subsetting and filtering mechanisms, and examine how these choices affect usability and performance for weather and climate applications. Together, these considerations motivate the need for more flexible and unified datacube structures that support both expressive representations, as well as efficient data access.
\subsection{Datacube Abstraction and Data Representations}
The datacube concept was first introduced in the field of business analytics as a way to organise and efficiently query multidimensional tabular databases \cite{datacube2}. Earlier datacube models supported operations such as aggregation, slicing and pivoting across different axes like product, region and time, providing a unified abstraction for structured multidimensional data \cite{datacube1}. As data-intensive scientific disciplines expanded, this abstraction proved broadly applicable beyond its original domain. Many fields, including biomedical imaging, genomics, environmental monitoring, social sciences and large-scale sensor networks, routinely generate datasets that can be described along multiple coordinate dimensions \cite{medical_datacube, social_datacube, network_datacube}. Over time, the datacube has therefore evolved into a widely adopted conceptual model for managing complex, high-dimensional datasets in a structured and interpretable form.\\
This evolution is reflected in scientific software frameworks such as xarray \cite{xarray} and storage formats such as Zarr \cite{zarr}, which represent datasets as labeled tensors or multidimensional arrays with named coordinates. These tensor-based models have become foundational in many data-driven workflows because they support intuitive indexing, efficient numerical computation and integration with high-performance computing and machine learning ecosystems \cite{pangeo_xarray, python_ml_xarray, xarray_scores}. For datasets that are dense, uniform and defined on orthogonal coordinate axes, such as regularly gridded simulations, standardised imaging data or homogeneous observational records, these representations provide both conceptual clarity and practical efficiency.\\
However, many real-world datasets deviate substantially from these assumptions. Across domains, data may be sparse, heterogeneous or defined only for particular combinations of variables. In medicine, patient records often contain missing values, condition-specific measurements or varying record lengths. Genomic datasets exhibit irregular coverage and alignment. Sensor network data may be intermittent or constrained by device-specific configurations. Meteorological and Earth system datasets, in particular, exhibit many of these challenges simultaneously. Climate simulations typically produce a single long time series, while weather forecasts introduce both an initialisation time and a forecast lead time, resulting in multiple temporal axes. Earth observation products add further dimensions related to instrument mode, orbit geometry or viewing configuration. As a result, no single dense, orthogonal tensor can faithfully represent the combined data space.\\
Within meteorological and climate datasets, further structural variability arises. Different variables may be available at different forecast lead times, vertical coordinates may be defined on model levels, pressure levels or only at the surface, and observational data may exist only under specific quality constraints or sensor configurations. This leads to branching data spaces in which dimensionality, coordinate definitions and data availability vary across subsets of the dataset. Although many subspaces share common structural patterns, the overall data space becomes irregular and sparse, with conditional dependencies that are difficult to express in traditional tensor models.\\
To address some of these limitations, xarray introduced DataTrees \cite{datatree1, datatree2}, a hierarchical abstraction in which related datacubes are organised into a tree structure. This allows groups of datasets to share common dimensions while permitting branching to represent subsets with different structures. Tree-based datacube models represent an important step toward capturing relational and conditionally defined data spaces, and they are particularly relevant for organising complex Earth system datasets composed of multiple related products.\\
Nevertheless, existing hierarchical datacube models still impose significant structural constraints. They typically assume fixed axis orderings, homogeneous dimensionality across sibling branches and a rigid notion of what constitutes a valid subspace. They struggle to represent datasets in which dimensionality itself varies across the domain, where different branches require different coordinate sets, or where successive constraints dynamically change the relevant dimensions. In such cases, the structure of the data cannot be expressed naturally without fragmentation or duplication. These limitations motivate the need for more general and flexible representations that can encode heterogeneous, sparse and structurally variable data spaces while preserving shared structure and enabling efficient navigation.
\subsection{Datacube Access and Data Extraction}
Beyond pure representation, an equally critical aspect of working with multidimensional datasets is how data is accessed and extracted from the underlying data stores. In practice, users rarely require entire datacubes. Instead, they seek specific subsets corresponding to regions of interest, time windows, vertical levels, ensemble members or combinations of conditional constraints. Effective datacube frameworks must therefore support data access strategies that allow users to retrieve only the information relevant to their scientific question, both efficiently and intuitively.\\
Traditional datacube frameworks support access through operations such as slicing, indexing and aggregation along coordinate axes \cite{datacube1, datacube_operations}. In tensor-based models, these operations are often tightly coupled to array indexing and chunking strategies \cite{chunking, array_indexing}, which work well for regular, dense datasets. However, as data spaces become more irregular and conditionally structured, these access patterns become increasingly difficult to generalise. Querying sparse or branching data often requires custom logic, preprocessing or manual traversal of multiple datacubes, reducing usability and increasing the computational cost of data access.\\
In response to these challenges, the concept of feature extraction has gained increasing prominence in meteorology and climate science as a higher-level data access strategy. Rather than exposing raw multidimensional arrays, feature extraction frameworks allow users to specify the subsets of data they are interested in, such as trajectories, cross-sections, vertical profiles or spatial regions, directly in terms of scientific features. This shift is reflected in the development of interfaces such as the OGC Environmental Data Retrieval (EDR) standard \cite{ogc-edr}, which enables users to request specific features from complex geospatial datasets through a unified API.\\
At the institutional level, similar ideas have emerged in systems such as the Polytope feature extraction framework at ECMWF \cite{polytope, polytope_pasc}, which allows users to describe complex extraction requests over high-dimensional weather and climate datasets. Related approaches are also present in data services built on platforms such as RasDaMan \cite{rasdaman1, rasdaman2} or the Sentinel Hub \cite{sentinel_hub}, where users can request spatial or temporal subsets through higher-level clipping and subsetting operations. These developments reflect the shift towards feature extraction as an essential way of interacting with modern datacubes and data systems.\\
However, many existing datacube implementations treat feature extraction as a post-processing step applied after data have already been retrieved from backend storage. In such systems, large volumes of data are first accessed and transferred before being clipped or post-processed to produce the final user output. Whilst this approach is often sufficient for smaller datasets, it becomes increasingly inefficient for weather and climate data, where backend archives can span petabytes and user requests typically target only a few megabytes of information. In such applications, the cost of accessing unnecessary data dominates the workflow, limiting scalability and performance.\\
This gap between datacube abstractions and efficient data access strategies points to the need for new datacube frameworks. In such frameworks, feature extraction should be treated as a key datacube operation and tightly integrated with the underlying data model. Access strategies should be able to exploit knowledge of the data structure itself, allowing selective traversal and retrieval of only the relevant subspaces. Bridging representation and feature-based data access in this way is essential for enabling scalable and user-friendly workflows in meteorology, climate science and other data-intensive disciplines.
\section{A Datacube Generalisation}
Traditional datacube models assume that data occupy a dense, orthogonal and fully defined multidimensional space. As shown in the previous sections, these assumptions increasingly break down for modern scientific datasets, particularly in meteorology and climate science, where data availability depends on conditional constraints, dimensionality varies across subsets and large portions of the space may be sparse or undefined. Representing such datasets as complete datacubes either leads to fragmentation into multiple disconnected cubes or to inefficient padding with missing values.\\
To address these limitations, we introduce a generalisation of the datacube concept that relaxes global assumptions of completeness and orthogonality. Instead of viewing a datacube as a single multidimensional array, we represent the data space as a compressed hierarchical structure that explicitly encodes branching and conditional relationships. This representation, which we refer to as a data hypercube, preserves shared structure where possible, while allowing divergence where required. It therefore provides a natural foundation for efficient querying and feature-oriented data access. In the remainder of this section, we describe this generalised datacube representation, formalise its structure and discuss the advantages it offers in terms of expressiveness, performance and reuse across meteorological data systems.
\subsection{Data Hypercubes} 
The generalised data hypercube represents multidimensional data spaces as compressed tree structures instead of dense datacubes. Each level of the tree corresponds to a dimension and branching represents conditional structure due to data constraints. Paths from the root to a leaf define valid combinations of coordinates for which data exists.\\
In practice, many geophysical datasets are characterised by constraints between dimensions, meaning that not all combinations of coordinates are meaningful or populated. For example, consider a meteorological dataset containing a set of atmospheric variables. Some variables, such as 2-metre temperature or surface pressure, are defined only at the surface, while others, such as temperature, wind or humidity, are defined on multiple pressure levels. The vertical dimension is therefore conditional on the variable, and treating all variables as sharing the same set of vertical coordinates would either require introducing artificial levels for surface-only variables or separating the dataset into multiple independent datacubes, thereby obscuring relationships between variables. Instead, the generalised data hypercube tree formulation allows dense, orthogonal sub-datacubes to be represented efficiently, while also accommodating sparse, heterogeneous and irregular data spaces.\\
Formally, a data hypercube can be defined as a rooted, directed tree
\[
\mathcal{T} = (V, E),
\]
where each node \(v \in V\) is associated with a dimension \(d(v)\) and a subset of admissible coordinate values \(C_v \subseteq \mathcal{C}_{d(v)}\). An edge \((v_i, v_j) \in E\) represents refinement by an additional constraint on a subsequent dimension. Each root-to-leaf path therefore defines a valid subspace of the overall domain, and leaf nodes can then be associated with payloads such as data arrays, storage references or various metadata. In this context, fully dense datacubes correspond to flat trees where each dimension level exhaustively enumerates its coordinate domain.\\
In the generalised representation, the structural variability in the example above is expressed through branching in the hypercube tree. Higher levels of the tree may encode shared dimensions such as forecast lead time, while a subsequent branching distinguishes between surface variables and pressure-level variables. One branch terminates at the surface, while the other introduces an additional vertical dimension corresponding to pressure levels. Each leaf node therefore represents an internally consistent subspace with a well-defined dimensional structure. Shared dimensions are represented once, and divergence is introduced only where the data structure genuinely differs.\\
Unlike classical datacubes, this representation is not invariant under permutation of dimensions. The order in which dimensions appear along the tree defines both the logical structure of the data space, as well as the optimal traversal order used for querying. Dimensions higher up in the tree act as early filters, which allows for pruning of entire subspaces when they are irrelevant to a given request. Selecting an appropriate dimension ordering therefore becomes an important design choice that can optimally be aligned with common access patterns. In meteorological and climate workflows, this may be implemented by placing higher-level distinctions, such as different satellite instruments or meteorological model types, towards the top of the tree, so that branches corresponding to incompatible acquisition systems or modelling frameworks can be excluded early in the traversal. Once the data space has been partitioned along these distinctions, the remaining dimensions within each resulting sub-hypercube are typically more homogeneous, allowing for more efficient querying and internal organisation within each branch.\\
This data hypercube representation provides a compact and expressive way to model complex meteorological data spaces. By explicitly encoding structural dependencies, it enables both efficient traversal as well as robust filtering, forming the basis for more advanced data extraction and analysis pipelines.\\
Tree-based datacube representations of this form have already been adopted in different operational contexts at ECMWF. They underpin the internal representation of data availability in the MARS archive \cite{mars} and are also used as the backend of the Destination Earth STAC catalogue \cite{qubed_catalogue} through the Qubed software \cite{ecmwf-qubed}. In these systems, data hypercubes have demonstrated their ability to efficiently encode large, heterogeneous meteorological data stores while supporting scalable data discovery and access.
\subsection{Data Hypercube Operations}
The data hypercube representations introduced previously are built on compressed tree structures that support a well-defined set of operations. These operations are essential both for constructing data hypercubes efficiently and for maintaining their optimality as the underlying data evolve over time. In practice, data hypercube trees must be built incrementally, recomputed as new data become available, and traversed efficiently in response to diverse user queries. A key part of this are the operations on the tree structure itself. Basic operations include traversal, filtering and slicing along dimensions, which enable selective exploration of the data space.\\
From a construction perspective, consider a data hypercube defined over a set of dimensions $D=\{d_1,\dots,d_k\}$, where each dimension $d_i$ has cardinality $n_i$. In practice, the resulting tree representation is not necessarily uniform. Indeed, due to sparsity or missing coordinate combinations, different branches of the tree may terminate at different tree depths. Let $k_{\max}$ denote the maximum depth of any root-to-leaf path in the tree. A naïve construction procedure inserts each available data tuple independently by creating a path through the tree corresponding to its coordinates. If the underlying dataset contains $N$ such tuples, which implies that $N$ is the number of leaf nodes, then each insertion requires at most traversing or creating a path of length $k_{\max}$. The total construction cost is therefore bounded by
\[
T_{\mathrm{construct}}^{\mathrm{naive}}=\mathcal{O}(k_{\max}N).
\]
Similarly, extraction or traversal operations in this representation may require visiting a large fraction of the $N$ leaf nodes, and set operations such as unions or intersections between two data hypercubes with $N_1$ and $N_2$ leaves require
\[
T_{\cup}^{\mathrm{naive}},\,T_{\cap}^{\mathrm{naive}}
=\mathcal{O}(N_1+N_2).
\]
More advanced set operations, such as unions and intersections of trees, are required to combine multiple datasets, reconcile overlapping data spaces or restrict a representation to shared subsets of dimensions and coordinates. These operations make it possible to construct complex data spaces successively from simpler ones. Union operations play a central role when integrating heterogeneous datasets or aggregating data over time, for example when successive forecast cycles or observational products are added to an existing data store. In such cases, multiple data hypercube trees must be merged into a single representation that preserves shared structure while introducing branching only where the data differs. In contrast, intersection operations serve the purpose of restricting a data hypercube to a common subspace, such as when identifying overlapping availability between datasets or applying constraints derived from user queries.\\
To ensure efficiency, these set operations must be accompanied by compression and re-optimisation of the tree structure. After unions or intersections, the resulting tree might contain redundant branches or uncompressed duplicated subtrees that degrade traversal performance. Let compression collapse structurally identical subtrees so that a tree with $N$ nodes is reduced to a representation with only
\[
M \ll N
\]
distinct structural nodes.\\
The compression operation itself consists of identifying structurally identical subtrees and replacing them with a shared representation. This requires comparing branches of the tree that span the same coordinate subspaces in order to detect similarities between sub-hypercubes. In the worst case, this comparison must be performed across all nodes of the uncompressed representation, so that the compression step is linear in the number of nodes,
\[
T_{\mathrm{compress}}=\mathcal{O}(N).
\]
In practice, however, compression is normally performed bottom-up, beginning at the leaves of the tree and propagating upwards. Since identical subtrees are more likely to occur towards the leaves of the tree, performing compression in this way reduces the number of distinct intermediate nodes that need to be compared higher up in the tree. As a result, tree hierarchies that maximise redundancy in deeper layers not only decrease the final number of structural nodes $M$, but also reduce the effective cost of the compression procedure itself.\\
After compression, subsequent construction and transformation operations can then be performed in terms of these unique sub-hypercubes rather than the full Cartesian space, yielding
\[
T_{\mathrm{construct}}^{\mathrm{compressed}}=\mathcal{O}(k_{\max}M),
\qquad
T_{\cup}^{\mathrm{compressed}},\,T_{\cap}^{\mathrm{compressed}}
=\mathcal{O}(M_1+M_2).
\]
Hence unions or intersections between compressed data hypercubes no longer require traversing all $N$ nodes, but only the $M$ distinct structural nodes, resulting in a reduction in computational cost by a factor proportional to $(N_1+N_2)/(M_1+M_2) \gg 1$.\\
Compression is therefore an essential operation as it collapses identical subtrees and factors out shared tree structure, restoring an efficient data hypercube compressed tree object after union, intersection or any other transformation. As a result, compression is an integral part of the data hypercube construction process. It ensures that the resulting representation remains compact and well aligned with typical access patterns. It also highlights that the achievable efficiency depends directly on how the dimensions are arranged in the tree hierarchy as dimension orderings that maximise shared sub-hypercubes minimise $M$, whereas early branching along highly variable dimensions increases $M$ and degrades the performance of construction, extraction and set operations.
\subsection{Data Hypercube Performance}
The performance of data hypercube representations is a central concern in any practical data workflow. Operations such as construction, compression, union and intersection are fundamental steps in building, updating and maintaining coherent representations of complex data spaces. As these operations are repeatedly applied throughout the lifetime of a data hypercube, their computational characteristics directly influence the scalability and performance of downstream applications and data access. Understanding how efficiently these operations can be performed is therefore essential for assessing the practical use of such generalised data hypercube frameworks.\\
Performance considerations are especially important in operational and time-critical settings, where data spaces evolve continuously and up-to-date representations must be generated and traversed with minimal latency. In such contexts, inefficiencies in hypercube construction or traversal can quickly start to dominate the overall system cost, limiting the ability to serve timely and consistent data products.
\subsubsection{Data Workflows}
In the following, we study performance using Qubes \cite{ecmwf-qubed} as a concrete implementation of compressed tree-based data hypercube representations. While the analysis focuses on Qubes, the results are representative of a broader class of generalised data hypercube frameworks that encode structure, sparsity and conditional relationships using hierarchical, compressed trees.\\
Within the workflows considered here, Qubes act primarily as fast indices and caches over large underlying data stores. Their effectiveness therefore depends not only on their expressive power, but also on how efficiently they can be constructed, maintained and reused as the underlying data change. Understanding the performance behaviour of core Qube operations is thus central to evaluating their suitability for operational use.\\
We analyse the performance of Qubes by considering their lifecycle within a typical data workflow. We first examine the cost of Qube construction, which represents the entry point to the data structure. We then study compression, focusing on its computational characteristics and its role in maintaining efficiency at scale. Compression introduces an inherent trade-off as, although it increases the cost of construction, it significantly reduces the cost of subsequent traversal, querying and set operations, which tend to dominate typical usage patterns. The analysis then extends to more complex manipulations, such as unions and other set operations, which are required when combining or incrementally updating data spaces. Finally, we illustrate these aspects through concrete timing results for the construction and manipulation of large-scale Earth system data hypercubes in weather forecasting and climate science, in the context of the Destination Earth initiative \cite{destine}.
\subsubsection{Qube Construction}
\begin{figure}[ht!]
\centering
    \subfigure[Increasing number of leaf nodes for a flat Qube.]{\includegraphics[width=0.45\columnwidth]{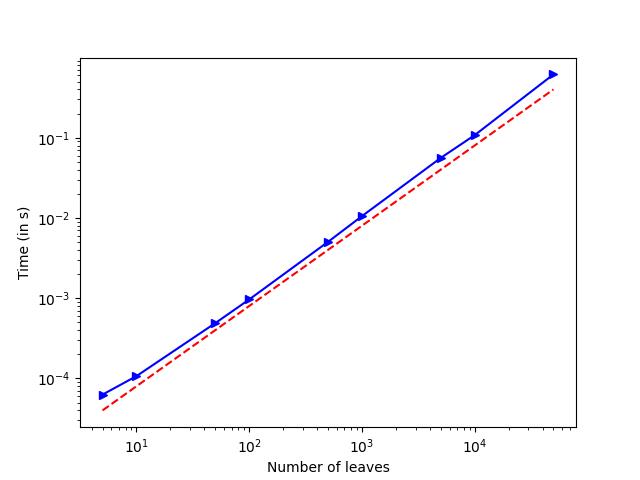}
    \label{construction_perf_1}}
    \hfill
 \subfigure[Branching at different Qube depths.]{\includegraphics[width=0.45\columnwidth]{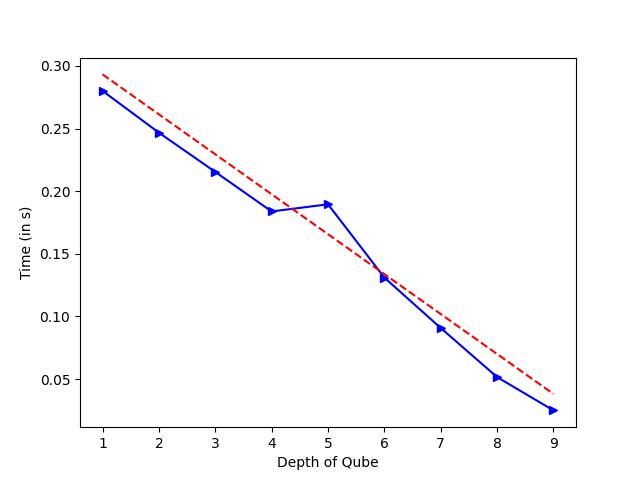}
 \label{construction_perf_2}}
 \hfill
 \subfigure[Increasing number of leaves for a wide Qube.]{\includegraphics[width=0.45\columnwidth]{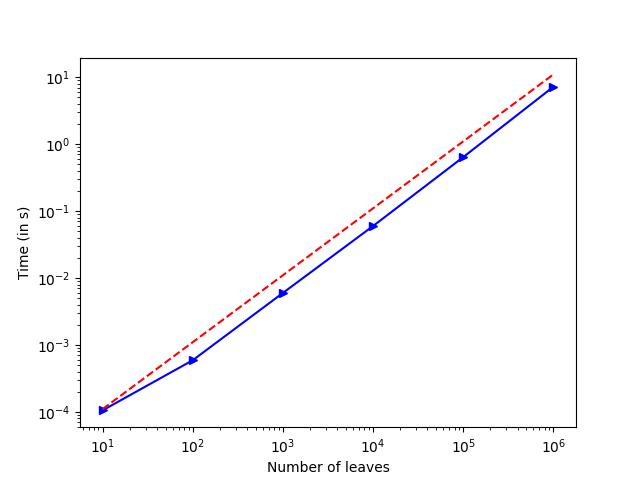}\label{construction_perf_3}}
 \caption{Time to construct different types of Qubes. The dashed lines represent the corresponding expected linear behaviour of these timings.}\label{construction_perf}
\end{figure}
The construction time of Qubes depends primarily on the tree size and on where branching occurs in the tree. In Figure \ref{construction_perf_1}, we first study the impact of the number of leaves by constructing a flat tree that branches only at the leaves and varying their count. As expected, the construction time increases linearly with the number of leaves, since this setup requires building a linearly growing number of nodes, which is the dominant cost.\\
In Figure \ref{construction_perf_2}, we next examine how the branching level affects construction time by comparing trees that branch at different depths. Branching closer to the root leads to higher construction times, while branching near the leaves is faster. This linear decrease in time is expected, because earlier branching produces more nodes overall.\\
Finally, Figure \ref{construction_perf_3} combines both effects by varying the number of nodes at each depth while maintaining a balanced tree. The construction time again grows linearly with the total number of nodes. However, because the number of nodes increases exponentially with the branching factor per level, even small increases in branching lead to large construction costs. Importantly, this cost is largely avoided for dense Qubes, where native compression reduces the effective number of leaves from $n$ to 1, substantially mitigating the overhead of highly branched trees.
\subsubsection{Compression}
\begin{figure}[ht!]
\centering
    \subfigure[Increasing number of leaf nodes for a flat Qube.]{\includegraphics[width=0.45\columnwidth]{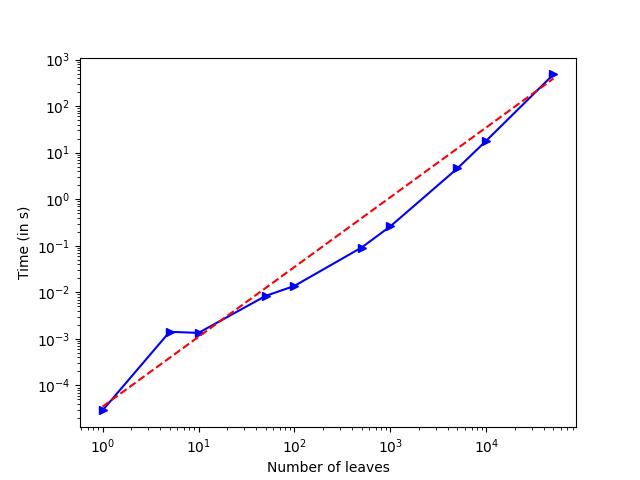}
    \label{compression_perf_1}}
    \hfill
 \subfigure[Branching at different Qube depths.]{\includegraphics[width=0.45\columnwidth]{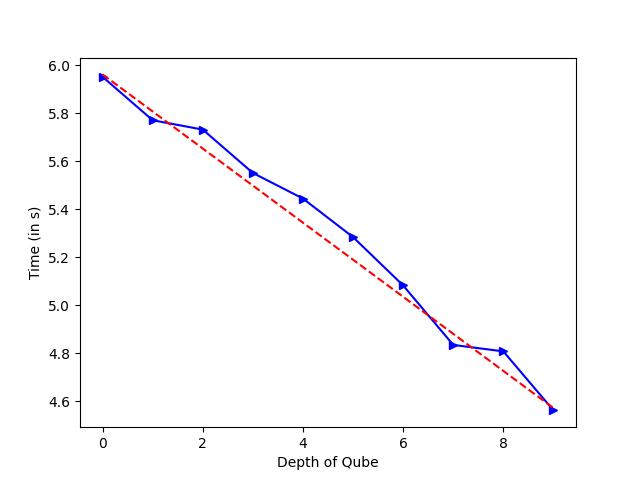}
 \label{compression_perf_2}}
 \hfill
 \subfigure[Increasing number of leaves for a wide Qube with high branching factor.]{\includegraphics[width=0.45\columnwidth]{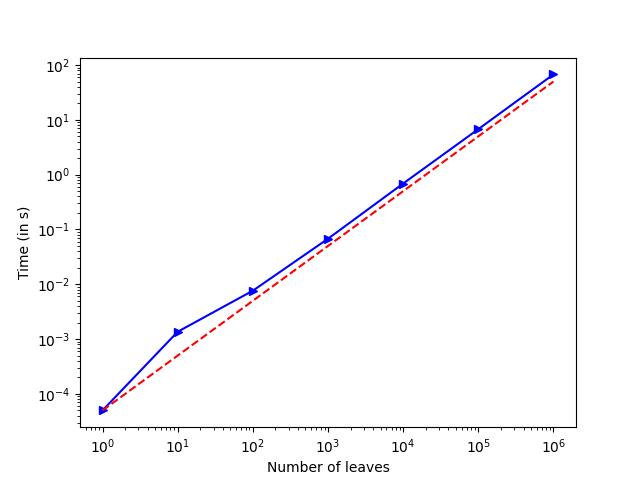}\label{compression_perf_3}}
 \caption{Time to compress different types of Qubes. The dashed lines represent the corresponding expected linear behaviour of these timings.}\label{compression_perf}
\end{figure}

Compression is a core feature of Qubes, as it enables a highly efficient representation of the data. We therefore evaluate the cost of compressing an initially uncompressed Qube. The experimental setup is identical to the previous section, but here we measure compression time rather than construction time.\\
The observed trends closely mirror those for construction. In Figure \ref{compression_perf_1}, the compression cost increases linearly with the number of leaf nodes, reflecting the need to process a linearly growing number of branches. In Figure \ref{compression_perf_2}, we further observe that branching higher in the tree is less efficient for compression, since detecting and merging equivalent subtrees requires deeper recursion. Nevertheless, this cost still scales linearly and as branching moves toward the leaves, fewer nested comparisons are required, making the compression progressively cheaper.\\
Figure \ref{compression_perf_3} combines these effects and again shows a linear relationship between compression time and the total number of leaves in the Qube. This behavior arises from the linear traversal of nested branches and the sequential merging of equivalent subtrees. Importantly, compression is significantly faster on already dense or partially compressed Qubes, as fewer branches need to be compared and merged.\\
In practice, compression is performed only once. After obtaining a maximally compressed Qube, subsequent operations benefit from this optimal representation and execute much more efficiently. Moreover, the absolute compression cost remains modest and is typically even lower in real-world settings, where Qubes tend to become denser toward the leaves, precisely where compression is most efficient as seen in Figure \ref{compression_perf_2}.
\subsubsection{Set Operations}
\begin{figure}[ht!]
\centering
    \subfigure[Union of two Qubes with an increasing number of total nodes in each.]{\includegraphics[width=0.45\columnwidth]{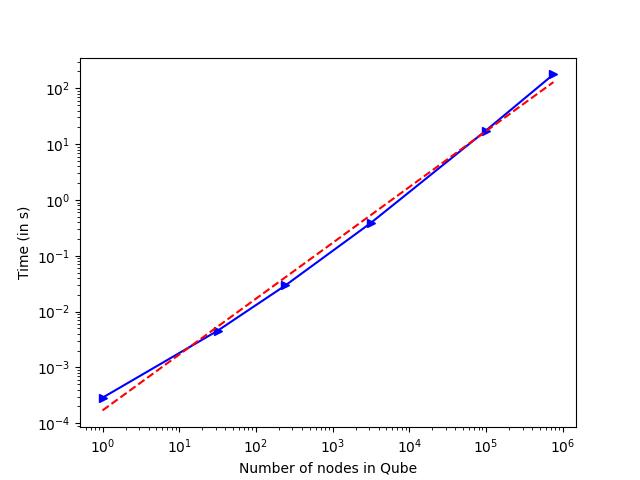}
    \label{union_perf_1}}
    \hfill
 \subfigure[Union of a growing number of medium-sized Qubes.]{\includegraphics[width=0.45\columnwidth]{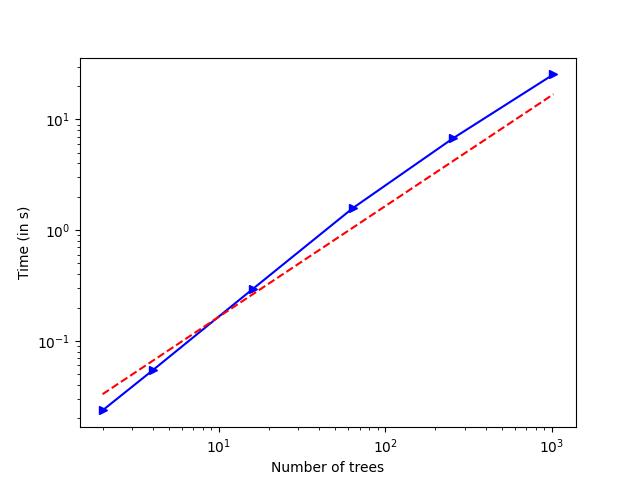}
 \label{union_perf_2}}
 \caption{Time to union Qubes together. The dashed lines represent the corresponding expected linear behaviour of these timings.}\label{union_perf}
\end{figure}

Once Qubes are constructed, the primary operations of interest are set operations. Since these operations exhibit similar performance characteristics, we focus on the union operation as a representative case. In Figure \ref{union_perf_1}, we first compare the cost of unioning fully expanded trees with that of highly compressed Qubes. The results show that union time increases linearly with the number of nodes per tree depth, confirming the theoretical bound that set operations scale linearly with the total number of nodes in the input Qubes.\\
In Figure \ref{union_perf_2}, we then study how performance scales when progressively unioning multiple Qubes of fixed size. The total cost again grows linearly, as each union is applied sequentially to the result of the previous one. This pattern is common in practice, particularly when constructing complex Qubes from several dense sub-Qubes. While the linear scaling is expected, it also suggests potential optimisations. Indeed, unioning more similar Qubes first or adopting alternative merge strategies that batch operations, could reduce the overall cost in more complex workflows.
\subsubsection{Practical Considerations}
In practice, constructing Qubes for real-world datasets requires combining several of the operations discussed above, including incremental construction, compression and merging. Overall performance therefore depends not only on the size of the data, but also on its structural properties and on the order in which operations are applied. While naive construction and merge operations typically scale linearly with the number of entries, this quickly becomes impractical for datasets containing millions of elements, making more informed construction strategies essential.\\
A key optimisation strategy is to exploit structural similarity between sub-hypercubes as early as possible in the construction process. By compressing smaller Qubes before merging them into larger structures, identical or near-identical subtrees can be collapsed early. This not only reduces memory usage, but also lowers the cost of subsequent operations. Early compression also limits the growth of intermediate representations and avoids repeated work on redundant structures.\\
Another effective approach is to batch merge operations together so that smaller and cheaper Qubes are combined first. This delays more expensive unions on large trees until sufficient structure has already been found and compressed in intermediate results. Similarly, compression and merging can be applied to lower levels of the Qube before the resulting branches are attached to higher-level tree nodes. By only performing expensive union operations when the data structure starts to diverge, we ensure that these transformations are applied to shallower subsets of the tree, which improves the overall efficiency.\\
These strategies are already used in practice within the Destination Earth digital twins, where Qubes act as fast, cache-like indexes of available data to support interactive catalogue exploration. To construct these indexes, we scan the flat metadata index exposed by ECMWF’s FDB data store \cite{fdb1, fdb2} and apply a combination of incremental construction, compression and merging to build a Qube that captures the structure of the available data. For the Climate digital twin \cite{climate_twin, climate_twin2}, which contains approximately 8.6 million data entries, Qube construction takes on the order of one day. In contrast, the smaller Extremes digital twin \cite{extremes_twin} can be regenerated daily in roughly one hour.\\
Despite these optimisations, Qube construction remains computationally expensive and should be viewed as a costly indexing step rather than a lightweight operation. Once built, however, Qubes provide a compact representation of the data structure and enable very fast selection and traversal operations. This makes them highly effective as query interfaces compared to direct access to the underlying data store. Overall, Qubes are best understood as slow-moving but fast index caches. They are expensive to generate, yet highly efficient to use and reuse within interactive and operational workflows.
\section{An Integrated Data Extraction System}
The tree-based data hypercube paradigm introduced in the previous section provides a compact and expressive representation of complex, sparse and irregular data spaces. We have discussed how structural constraints and conditional data availability can be encoded directly in this data representation itself, and have shown, using Qubes as a concrete instantiation of this paradigm, how this representation can be efficiently constructed and reused.\\
In this section, we build on these foundations and present a complete feature extraction system that integrates structure-aware data hypercube representations directly into an operational data access pipeline. Rather than treating data hypercubes as passive data containers, the system incorporates data structure, indexing and access strategies into a single coherent workflow. This allows feature extraction algorithms to reason about data availability and layout directly from the data hypercube representation, without relying on static, externally maintained rules about the data.\\
This section is organised as follows. We first describe the overall system architecture and the interaction between the Polytope \cite{polytope_software}, Qubed \cite{ecmwf-qubed} and GribJump \cite{gribjump} software, highlighting how structure-aware data hypercube representations are incorporated into the extraction workflow. We then analyse the performance characteristics of the system, with particular emphasis on I/O efficiency and how tree-based representations influence access patterns. Finally, we discuss the implications of this approach for user interaction, showing how it enables new, more expressive access patterns and allows users to request precisely defined scientific features from complex datasets without detailed knowledge of data layout or storage constraints.
\subsection{System Architecture}
\begin{figure}[ht!]
\centering
\includegraphics[width=\columnwidth]{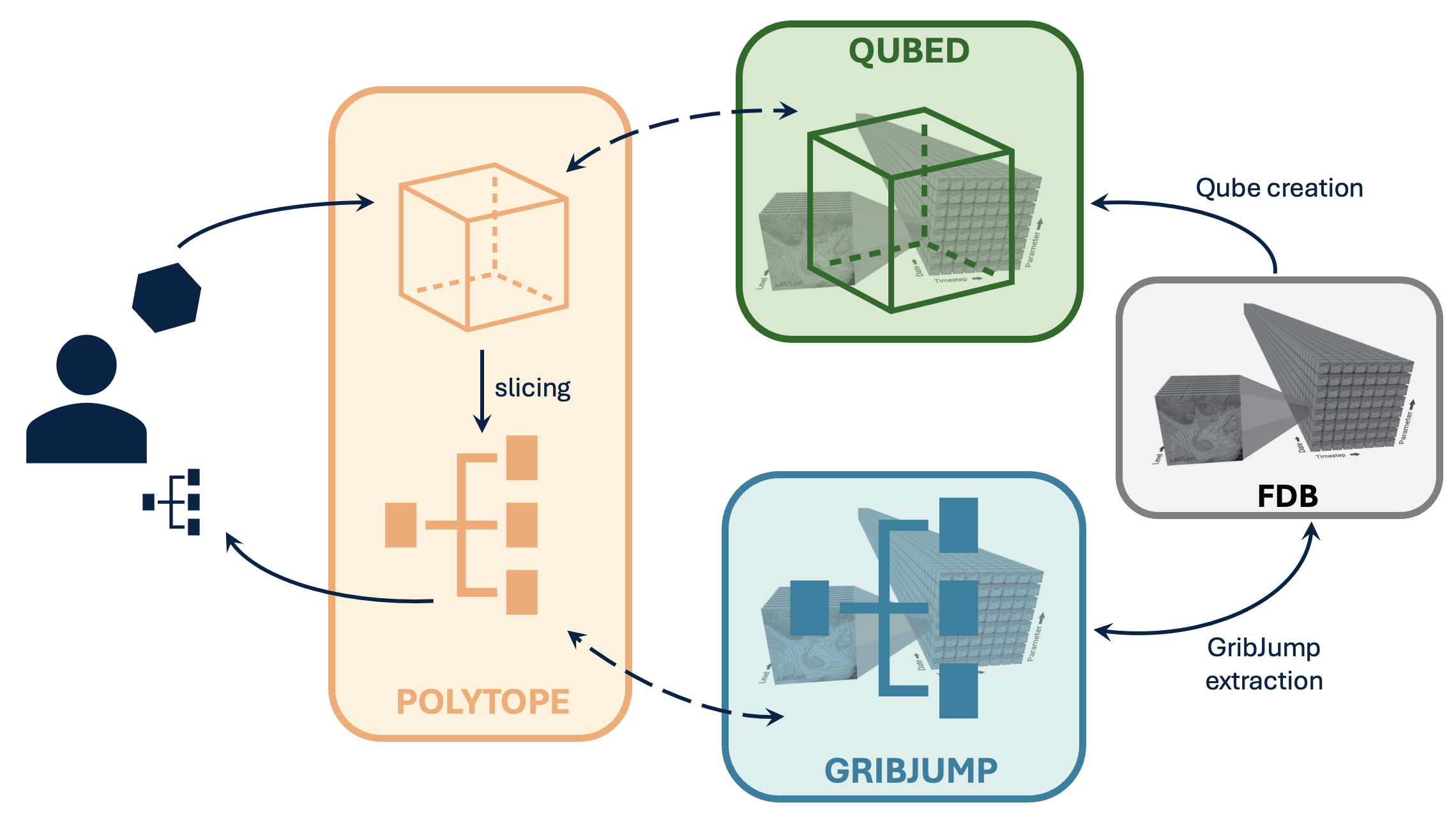}
 \caption{Full feature extraction system with the Polytope, GribJump and Qubed components.}\label{system}
\end{figure}

The integrated feature extraction system is organised around a generic, tree-based data hypercube representation that provides a uniform and faithful view of the available data. All interactions between user requests and the underlying data stores are mediated through this data hypercube tree, which encodes the actual structure of the data space rather than an idealised dense abstraction. In our implementation, this representation is realised using Qubes. Sparsity, structural constraints and branching are encoded explicitly, allowing the system to reason directly about which combinations of coordinates exist without relying on external configuration rules. \\
The resulting system combines three tightly integrated components, Polytope, which provides the geometric and algorithmic core for feature extraction, Qubed, which maintains a compact, tree-based index of the available data, and GribJump, which enables fine-grained, byte-level access to the underlying storage backends. Together, these components transform feature extraction from a field- or chunk-oriented operation into a structured traversal and filtering process over a precise representation of the data space.\\
Qubes are constructed and maintained by the Qubed software \cite{ecmwf-qubed}. Qubed builds this representation by scanning the flat metadata index exposed by the FDB data store \cite{fdb1, fdb2} and organising the available keys into a compressed tree structure. Each node in the tree corresponds to a pair of dimension and associated coordinates, and branches represent conditional structure within the dataset, such as dimensions that are only present for specific models, resolutions or forecast steps. Because dataset structure typically evolves slowly, the resulting Qube can be constructed as part of a pre-computation phase and cached for reuse across many extraction requests.\\
Feature extraction is performed by Polytope \cite{polytope}, which operates directly on this Qube representation. User requests, expressed in terms of scientific features such as points, trajectories, regions or timeseries, are translated by Polytope into a set of constraints over the abstract data space. Rather than constructing a dense, orthogonal view of the data hypercube by taking Cartesian products of coordinate values, Polytope traverses the Qube tree and incrementally filters it by pruning branches that are incompatible with the request. This process both ensures that the system accesses only valid combinations of coordinates in the dataset, but also that all structural constraints encoded in the data are respected throughout the extraction workflow.\\
The filtering operation produces a reduced Qube, which shows the exact data subsets to extract from the backend stores for a given user request. For each leaf in this resulting tree, Polytope then computes the corresponding data store indices. These are forwarded to GribJump \cite{gribjump}, which implements the backend data access layer by providing byte-level access to data contained inside of FDB data stores. Instead of loading complete GRIB fields, or layers of the atmosphere, it retrieves only the byte ranges required to satisfy the request. This both reduces system I/O and eliminates the need for post-extraction clipping. The access model remains fully compatible with existing storage backends and file formats, as it operates directly on the underlying data representations without requiring new storage layouts.\\
A key architectural principle of the system is the explicit decoupling of logical data organisation from physical storage. The Qube representation captures logical structure, availability and constraints, while GribJump handles physical access to stored data. Polytope acts as the coordinating layer that translates between these two domains, ensuring that extraction logic is driven by data structure rather than by assumptions about the underlying storage layout.\\
Compared to existing multidimensional data frameworks such as xarray \cite{xarray}, Zarr \cite{zarr} or RasDaMan \cite{rasdaman1, rasdaman2}, which typically rely on dense arrays, predefined chunking strategies or rigid data models, this architecture generalises the notion of a datacube to encompass sparse, heterogeneous and conditionally defined data spaces. By combining tree-based data hypercubes, geometric filtering algorithms as well as byte-level data access into a single framework, the system provides a flexible foundation for efficient feature extraction across a wide range of meteorological and climate datasets.\\
The complete extraction workflow is illustrated in Figure \ref{system}, where a Qube index is first constructed once from the FDB metadata and cached. User-defined extraction requests are then processed by Polytope through structured traversal and filtering of this index. Finally, the resulting set of valid data indices is then passed to GribJump, which retrieves only the required bytes from the data store to assemble the final extracted feature.
\subsection{System Performance}
Many of the performance evaluations of the Polytope–GribJump system presented in earlier work \cite{polytope_pasc} continue to apply to the new extraction pipeline. An important difference, however, is that the expensive pre-computation phase previously required to construct a datacube representation during extraction is now bypassed. Instead, the system reads an existing Qube representation, which acts as a cached view of the data space. This significantly reduces performance overhead, as the pre-computation phase is now almost instantaneous, and is particularly beneficial to enable true interactive workflows.\\
The introduction of the Qubed abstraction adds only minimal runtime overhead because it serves as a read-only cache queried during extraction rather than as an active processing component. As with all data access systems, however, overall performance remains closely tied to access patterns and to how well these patterns align with the underlying storage layout. When the tree structure reflects how data are physically stored, it naturally highlights access paths that group nearby data, enabling more efficient retrieval and offering guidance on optimal usage patterns.\\
In practice, requests aligned with native storage layouts, such as spatial subsets within a single field, are served more quickly than requests spanning many fields. Polytope mitigates this effect by minimising total I/O and grouping access to consecutively stored data wherever possible, although algorithmic cost still increases with the size and geometric complexity of the requested feature. These effects are clearly visible in typical usage. Extracting a single full field using native access methods often completes in fractions of a second, whereas Polytope-based extraction may take a few seconds due to additional geometric processing. However, Polytope returns only a small fraction of the data, which is already advantageous in many scenarios.\\
As extraction scales across many fields, the benefits become more pronounced. For example, extracting a full forecast timeseries spanning 96 fields requires several seconds when using traditional full-field access, while the Polytope-based approach retrieves only a single point per field and scales minimally. When extending to ensemble forecasts involving hundreds of fields, traditional extraction can take minutes, whereas feature extraction remains on the order of seconds. This demonstrates the effectiveness of the approach for access patterns that are common in practice but poorly supported by native data layouts.\\
At the same time, performance remains influenced by the underlying storage system and hardware configuration. Factors such as data ordering, compression schemes, caching behaviour and storage media performance cannot be fully abstracted away. As a result, deploying the framework on a new backend or platform requires careful inspection of access patterns and dedicated performance analysis. This dependency is not unique to the proposed system and applies equally to other widely used data access frameworks, including xarray- and Zarr-based workflows. In all such systems, performance can degrade when user access patterns diverge from assumptions made during data layout and indexing.\\
Users should therefore be aware that extraction performance is determined not only by the feature extraction algorithm itself, but also by how data are organised and accessed at the storage level. While the framework provides strong guarantees on minimising unnecessary data access, optimal performance ultimately requires an overall view of the full data access stack, from user request patterns down to the physical storage layer.
\subsection{User-Centric Workflows and New Access Patterns}
Beyond performance, a central strength of the proposed feature extraction framework lies in how it fundamentally changes the way users interact with large-scale weather and climate datasets. Traditional access mechanisms in NWP and climate data stores typically expose data at the level of complete fields or atmospheric layers. As a result, users are often forced to retrieve large volumes of data even when only a small subset is required, placing a significant burden on local storage, data transfer and post-processing before any scientific analysis can begin.\\
The framework developed here addresses this limitation by shifting feature extraction from a post-processing step to a core operation within the data access layer itself. The compressed tree data hypercube abstraction plays an important role in this transition. By explicitly encoding which data actually exists, along with its intrinsic constraints and branching structure, the tree view enables users to navigate and query the data space safely and naturally. Requests are guaranteed to only access existing data, as the extraction system provides strong semantic guarantees about the validity of user queries.\\
From a user perspective, this enables a more intuitive and application-oriented interaction model. Rather than formulating requests in terms of storage-specific concepts such as files, fields or chunks, users can express their needs directly in terms of scientific features of interest, including point time series, trajectories, regions or arbitrary spatio-temporal subsets. These high-level requests are translated internally by Polytope into precise byte-level access patterns, using the Qubed representation to reason about data availability and structure. As a result, users receive only the exact data that they need, in a format which is directly useful for downstream analysis.\\
This design significantly reduces the effort required to work with meteorological datasets. Indeed, users no longer need detailed knowledge of data formats, grid definitions or backend storage conventions in order to retrieve meteorological data efficiently. They can instead focus on scientific questions, whilst the extraction pipeline transparently manages the complexity of identifying, filtering and accessing the relevant data. This is particularly beneficial for non-expert users, interdisciplinary researchers and application developers, who rely on weather and climate data as inputs rather than as primary research objects.\\
The advantages of this approach are especially apparent in interactive and exploratory workflows. Because feature extraction retrieves only small targeted subsets of data, requests can often be served quickly enough to support iterative analysis, visualisation and prototyping. Users can explore timeseries at multiple locations, compare ensemble members or refine regions of interest without repeatedly downloading large datasets. Such interactivity is difficult to achieve with traditional full-field access and enables new classes of workflows, including web-based applications, notebooks and real-time decision-support systems.\\
The framework also improves accessibility by reducing local resource requirements. Full-resolution fields from modern climate simulations can easily exceed the memory and storage capacity of standard user environments. Feature-based access, by contrast, returns compact datasets that can be processed on modest local hardware or within lightweight cloud-based environments. This broadens access to large-scale weather and climate data while also supporting more sustainable data usage by minimising unnecessary data movement and duplication.\\
Finally, the generic nature of the feature extraction service allows it to operate across multiple data systems in a consistent manner. By separating abstract data representation from physical storage, the framework can serve as a common access layer over heterogeneous backends. This makes it well suited as a backend for standardised data access interfaces such as the OGC Environmental Data Retrieval (EDR) API \cite{ogc-edr} and enables seamless integration with data catalogues \cite{qubed_catalogue}, notification services \cite{aviso} and downstream processing pipelines \cite{earthkit}. In operational contexts such as the Destination Earth digital twins \cite{destine}, this unified access model allows users to discover, request and consume data through a single coherent interface, even when the underlying data are distributed across multiple systems.\\
Overall, the framework shifts the emphasis of data access from bulk data movement to information delivery. By allowing users to request precisely the data they need, in a form that is immediately usable, it enables more efficient, accessible and user-centred interaction with large scientific data stores. This perspective is central to the operational value of the system and highlights its potential as a foundational component for future data access services in weather, climate and Earth system science.
\section{Discussion and Future Work}
Feature extraction is becoming an increasingly important and widely used data access pattern in weather and climate science. As datasets continue to grow in size and complexity, users are progressively shifting away from full-field access towards more selective, application-driven queries that retrieve only the data strictly required for analysis. This trend is clearly visible in operational systems such as the Copernicus Data Store (CDS) \cite{cds}, where a substantial fraction of users request spatial or spatio-temporal subsets rather than complete fields. Supporting such access patterns efficiently and robustly is therefore a key requirement for modern data services.\\
In this work, we have presented a complete and fully generic feature extraction framework that operates on arbitrary data hypercube representations by reasoning directly over a tree-based abstraction of the data space. By decoupling logical data organisation from physical storage and by avoiding assumptions about data completeness, layout or regularity, the framework generalises traditional datacube models to encompass sparse, heterogeneous and conditionally defined datasets. Feature extraction is no longer treated as a post-processing step applied after data retrieval, but as a core operation embedded directly into the representation and traversal of the data space.\\
A central strength of the proposed approach is its generality. All data- and storage-specific details are encapsulated behind generic abstractions, allowing the same extraction pipeline to be reused across different datasets, storage systems and operational contexts. This makes the framework particularly well adapted to environments in which various data sources with different formats and conventions coexist, and where user access patterns vary widely. The ability to redeploy the system with minimal changes across different platforms is a key advantage as data infrastructures continue to evolve.\\
The practical relevance of this approach is reinforced by observed user behaviour in operational services. In the CDS, approximately 62\% of users accessing the ERA5 single-levels dataset apply an area constraint, with around 22\% targeting regions smaller than 10° × 10°. These access patterns are naturally well aligned with feature extraction and stand to benefit directly from reduced data transfer and backend I/O. While the precise performance gains depend on hardware and storage configuration, the framework therefore has clear potential to alleviate I/O bottlenecks and improve scalability across the Copernicus ecosystem.\\
This potential is already demonstrated by the operational deployment of the Polytope-based extraction service within the Destination Earth initiative \cite{destine, destine2}, where it is used daily to access Weather and Climate Digital Twin data. In this setting, feature extraction acts as the primary data access mechanism, supporting interactive workflows while integrating seamlessly with existing catalogues and services. Building on this experience, the framework is planned for integration into the unified CDS, where it is expected to benefit a large fraction of users by providing faster, more efficient and more accessible data extraction.\\
Several directions for future work emerge from this study. A first priority is the integration and evaluation of the framework across a broader range of backend data stores and hardware platforms. While the system is designed to minimise I/O and to align access patterns with underlying storage layouts, overall performance inevitably depends on backend-specific characteristics. Systematic evaluation across different storage technologies will therefore be essential to understand trade-offs and to guide deployment strategies.\\
A second direction concerns further optimisation based on observed user access patterns. By analysing how users interact with data in practice, it may be possible to adapt data hypercube representations, traversal orders or caching strategies to better match dominant workflows. This opens the door to access pattern optimisations that go beyond generic performance tuning.\\
Finally, the framework naturally lends itself to the integration of richer metadata into the data hypercube representation. Embedding additional information, such as grid definitions, physical data location, routing constraints or service-specific annotations, directly into the tree-based abstraction would allow the extraction system to make more informed decisions and to operate in a fully information-driven manner. Such metadata-enriched data hypercubes would further reduce the need for hard-coded assumptions and enable truly generic, self-describing data access across heterogeneous systems.\\
Overall, this work demonstrates how the combination of compressed tree-based data hypercube representations, geometric feature extraction and byte-level data access results in a coherent, scalable and user-centric data access framework. By bridging the gap between backend storage systems and user-driven workflows, the approach provides a practical foundation for future data services in weather, climate and Earth system science, where efficient, flexible and sustainable access to large datasets is increasingly essential.
\section*{Acknowledgements}
    The work presented in this paper has been produced in the context of the European Union’s Destination Earth Initiative and relates to tasks entrusted by the European Union to the European Centre for Medium-Range Weather Forecasts implementing part of this Initiative with funding by the European Union. Views and opinions expressed are those of the author(s) only and do not necessarily reflect those of the European Union or the European Commission. Neither the European Union nor the European Commission can be held responsible for them. 
\bibliography{biblio}

@inproceedings{polytope_pasc,
  title={Performance Analysis of an Efficient Algorithm for Feature Extraction from Large Scale Meteorological Data Stores},
  author={Leuridan, Mathilde and Bradley, Christopher and Hawkes, James and Quintino, Tiago and Schultz, Martin},
  booktitle={Proceedings of the Platform for Advanced Scientific Computing Conference},
  pages={1--9},
  year={2025}
}

@article{netcdf,
  title={NetCDF: an interface for scientific data access},
  author={Rew, Russ and Davis, Glenn},
  journal={IEEE computer graphics and applications},
  volume={10},
  number={4},
  pages={76--82},
  year={1990},
  publisher={IEEE}
}

@article{cmor,
  title={The Climate Model Output Rewriter (CMOR)},
  author={Mauzey, Chris and Doutriaux, Charles and Nadeau, Denis and Taylor, Karl E and Durack, Paul J and Betts, Edward and Cofino, Antonio S and Florek, Piotr and Hogan, Emma and Rodriguez Gonzalez, Jose M and others},
  journal={Zenodo},
  year={2025}
}

@misc{cf,
  title={NetCDF Climate and Forecast (CF) metadata conventions},
  author={Eaton, Brian and Gregory, Jonathan and Drach, Bob and Taylor, Karl and Hankin, Steve and Caron, John and Signell, Rich and Bentley, Phil and Rappa, Greg and H{\"o}ck, Heinke and others},
  year={2003},
  publisher={Version}
}

@misc{earthkit,
  title={Introducing earthkit},
  author={Russell, Iain and Quintino, Tiago and Raoult, Baudouin and Kert{\'e}sz, S{\'a}ndor and Maciel, Pedro and Varndell, James and de Wiart, Corentin Carton and Comyn-Platt, Edward and Iffrig, Olivier and Hawkes, James and others},
  year={2024}
}

@inproceedings{aviso,
  title={Aviso: Bridging HPC and Cloud with High-Throughput Notification System for NWP Data Availability},
  author={Iacopino, Claudio and Hawkes, James N and Quintino, Tiago and Raoult, Baudouin},
  booktitle={American Meteorological Society Meeting Abstracts},
  volume={101},
  pages={10--8},
  year={2021}
}

@techreport{cds,
  title={A modernised Data Store infrastructure for improving the access to Copernicus Climate and Atmosphere data and services.},
  author={Lopez, Angel and Buontempo, Carlo and Suttie, Martin and Raoult, Baudouin and Comyn-Platt, Edward and Varndell, James},
  year={2022},
  institution={Copernicus Meetings}
}

@article{sentinel_hub,
  title={Report of new features of the Sentinel Hub App},
  author={Lorilla, Roxanne Suzette},
  journal={Zenodo (CERN European Organization for Nuclear Research)},
  year={2025}
}

@article{ogc-edr,
  title={{OGC API-Environmental Data Retrieval Standard}},
  author={Burgoyne, M and Blodgett, D and Heazel, C and Little, C},
  journal={Open Geospatial Consortium Inc., Wayland, MA, USA, OpenGIS{\textregistered} Implementation Specification OGC}
}

@misc{ecmwf-qubed,
  title        = {Qubed},
  author       = {{European Centre for Medium-Range Weather Forecasts (ECMWF)}},
  year         = {2025},
  howpublished = {\url{https://github.com/ecmwf/qubed}},
  note         = {Accessed on 2026-01-05.},
}

@article{destine2,
  title={Digital twins, the journey of an operational weather prediction system into the heart of Destination Earth},
  author={Geenen, Thomas and Wedi, Nils and Milinski, Sebastian and Hadade, Ioan and Reuter, Balthasar and Smart, Simon and Hawkes, James and Kuwertz, Emma and Quintino, Tiago and Danovaro, Emanuele and others},
  journal={Procedia Computer Science},
  volume={240},
  pages={99--109},
  year={2024},
  publisher={Elsevier}
}

@article{array_indexing,
  title={A spatiotemporal indexing approach for efficient processing of big array-based climate data with MapReduce},
  author={Li, Zhenlong and Hu, Fei and Schnase, John L and Duffy, Daniel Q and Lee, Tsengdar and Bowen, Michael K and Yang, Chaowei},
  journal={International Journal of Geographical Information Science},
  volume={31},
  number={1},
  pages={17--35},
  year={2017},
  publisher={Taylor \& Francis}
}

@inproceedings{chunking,
  title={Optimal chunking of large multidimensional arrays for data warehousing},
  author={Otoo, Ekow J and Rotem, Doron and Seshadri, Sridhar},
  booktitle={Proceedings of the ACM tenth international workshop on Data warehousing and OLAP},
  pages={25--32},
  year={2007}
}

@article{mars,
  title={MARS, ECMWF's Meteorological Archive: Experience in managing a large archive},
  author={Raoult, Baudouin},
  year={2013}
}

@misc{gribjump,
  author       = {{ECMWF}},
  title        = {{GribJump}},
  howpublished = {\url{https://github.com/ecmwf/gribjump}},
  note         = {Accessed on 2026-02-04},
  year         = {2025},
  organization = {European Centre for Medium-Range Weather Forecasts},
  url          = {https://github.com/ecmwf/gribjump}
}

@misc{qubed_catalogue,
  title        = {{Qubed Catalogue Browser}},
  howpublished = {\url{https://qubed.lumi.apps.dte.destination-earth.eu/}},
  note         = {Accessed on 2026-02-04},
  year         = {2025},
  organization = {European Centre for Medium-Range Weather Forecasts},
  url          = {https://qubed.lumi.apps.dte.destination-earth.eu/}
}

@article{polytope,
  title={Polytope: an algorithm for efficient feature extraction on hypercubes},
  author={Leuridan, Mathilde and Hawkes, James and Smart, Simon and Danovaro, Emanuele and Schultz, Martin and Quintino, Tiago},
  journal={Journal of Big Data},
  volume={12},
  number={1},
  pages={1--25},
  year={2025},
  publisher={Springer}
}

@software{polytope_software,
  author       = {Mathilde Leuridan and
                  Adam Warde and
                  Oisin-M and
                  James Hawkes and
                  James Varndell and
                  Peter Tsrunchev and
                  Chris Bradley and
                  Dušan Figala and
                  Iain Russell and
                  Kai Kratz and
                  Simon Smart and
                  Yordan Radev},
  title        = {ecmwf/polytope: 2.1.1},
  month        = oct,
  year         = 2025,
  publisher    = {Zenodo},
  version      = {2.1.1},
  doi          = {10.5281/zenodo.17483028},
  url          = {https://doi.org/10.5281/zenodo.17483028},
  swhid        = {swh:1:dir:00d1c6957b80e5f9daecf6c95c65f88f56e9bacb
                   ;origin=https://doi.org/10.5281/zenodo.14537048;vi
                   sit=swh:1:snp:5dc53e5705a838d117b8c692fb6e881379db
                   386c;anchor=swh:1:rel:e2126c9be11f995620955b84df12
                   7e9b7d34e4b1;path=ecmwf-polytope-cb564b8
                  },
}

@misc{climate_twin,
  title={Destination Earth--A digital twin in support of climate services},
  author={Hoffmann, J{\"o}rn and Bauer, Peter and Sandu, Irina and Wedi, Nils and Geenen, Thomas and Thiemert, Daniel},
  year={2023},
  publisher={Elsevier}
}

@article{climate_twin2,
  title={The Destination Earth digital twin for climate change adaptation},
  author={Doblas-Reyes, Francisco J and Kontkanen, Jenni and Sandu, Irina and Acosta, Mario and Al Turjmam, Mohammed Hussam and Alsina-Ferrer, Ivan and Andr{\'e}s-Mart{\'\i}nez, Miguel and Arriola, Leo and Axness, Marvin and Batlle Mart{\'\i}n, Marc and others},
  journal={EGUsphere},
  volume={2025},
  pages={1--41},
  year={2025},
  publisher={Copernicus Publications G{\"o}ttingen, Germany}
}

@techreport{extremes_twin,
  title={Advances towards a better prediction of weather extremes in the Destination Earth initiative},
  author={Gasc{\'o}n, Est{\'\i}baliz and Sandu, Irina and Vanni{\`e}re, Beno{\^\i}t and Magnusson, Linus and Forbes, Richard and Polichtchouk, Inna and Van Niekerk, Annelize and S{\"u}tzl, Birgit and Maier-Gerber, Michael and Diamantakis, Michail and others},
  year={2023},
  institution={Copernicus Meetings}
}

@inproceedings{fdb1,
  title={A scalable object store for meteorological and climate data},
  author={Smart, Simon D and Quintino, Tiago and Raoult, Baudouin},
  booktitle={Proceedings of the Platform for Advanced Scientific Computing Conference},
  pages={1--8},
  year={2017}
}

@inproceedings{fdb2,
  title={A high-performance distributed object-store for exascale numerical weather prediction and climate},
  author={Smart, Simon D and Quintino, Tiago and Raoult, Baudouin},
  booktitle={Proceedings of the Platform for Advanced Scientific Computing Conference},
  pages={1--11},
  year={2019}
}

@article{big_data_weather,
  title={Big data analytics in weather forecasting: A systematic review},
  author={Fathi, Marzieh and Haghi Kashani, Mostafa and Jameii, Seyed Mahdi and Mahdipour, Ebrahim},
  journal={Archives of Computational Methods in Engineering},
  volume={29},
  number={2},
  pages={1247--1275},
  year={2022},
  publisher={Springer}
}

@article{big_data_climate,
  title={Climate data challenges in the 21st century},
  author={Overpeck, Jonathan T and Meehl, Gerald A and Bony, Sandrine and Easterling, David R},
  journal={science},
  volume={331},
  number={6018},
  pages={700--702},
  year={2011},
  publisher={American Association for the Advancement of Science}
}

@article{big_data_eo1,
  title={Earth observation big data for climate change research},
  author={Guo, Hua-Dong and Zhang, Li and Zhu, Lan-Wei},
  journal={Advances in Climate Change Research},
  volume={6},
  number={2},
  pages={108--117},
  year={2015},
  publisher={Elsevier}
}

@article{xarray,
  title={xarray: ND labeled arrays and datasets in Python},
  author={Hoyer, Stephan and Hamman, Joe},
  journal={Journal of open research software},
  volume={5},
  number={1},
  pages={10--10},
  year={2017}
}

@article{xarray_scores,
  title={scores: A Python package for verifying and evaluating models and predictions with xarray},
  author={Leeuwenburg, Tennessee and Loveday, Nicholas and Ebert, Elizabeth E and Cook, Harrison and Khanarmuei, Mohammadreza and Taggart, Robert J and Ramanathan, Nikeeth and Carroll, Maree and Chong, Stephanie and Griffiths, Aidan and others},
  journal={arXiv preprint arXiv:2406.07817},
  year={2024}
}

@book{python_ml_xarray,
  title={Python for probability, statistics, and machine learning},
  author={Unpingco, Jos{\'e}},
  volume={1},
  year={2016},
  publisher={Springer}
}

@inproceedings{pangeo_xarray,
  title={{The PANGEO Big Data Ecosystem and its use at CNES}},
  author={Eynard-Bontemps, Guillaume and Abernathey, Ryan and Hamman, Joseph and Ponte, Aurelien and Rath, Willi},
  booktitle={Big Data from Space (BiDS'19).... Turning Data into insights... 19-21 f{\'e}bruary 2019, Munich, Germany},
  year={2019}
}

@misc{datatree1,
  title        = {xarray-contrib/datatree: Prototype implementation of a tree-like hierarchical data structure for xarray},
  author       = {{xarray-contrib}},
  howpublished = {\url{https://github.com/xarray-contrib/datatree}},
  note         = {GitHub repository (archived). Accessed on 2025-12-30. },
  year         = {2024},
  url          = {https://github.com/xarray-contrib/datatree},
}

@inproceedings{datatree2,
  title={{Xarray-Datatree: Hierarchical Data Structures for Multi-Model Science}},
  author={Nicholas, Thomas and Busecke, Julius},
  booktitle={103rd AMS Annual Meeting},
  year={2023},
  organization={AMS}
}

@misc{zarr,
  title        = {{Zarr: Chunked, Compressed, N-Dimensional Arrays}},
  author       = {{Zarr Developers}},
  howpublished = {\url{https://zarr.dev/}},
  note         = {Accessed on 2025-12-30},
  year         = {2025},
  url          = {https://zarr.dev/},
}

@article{destine,
  title={Destination earth: High-performance computing for weather and climate},
  author={Wedi, Nils and Bauer, Peter and Sandu, Irina and Hoffmann, J{\"o}rn and Sheridan, Sophia and Cereceda, Rafael and Quintino, Tiago and Thiemert, Daniel and Geenen, Thomas},
  journal={Computing in Science \& Engineering},
  volume={24},
  number={6},
  pages={29--37},
  year={2023},
  publisher={IEEE}
}

@inproceedings{datacube_operations,
  title={Process cubes: Slicing, dicing, rolling up and drilling down event data for process mining},
  author={Van Der Aalst, Wil MP},
  booktitle={Asia-Pacific conference on business process management},
  pages={1--22},
  year={2013},
  organization={Springer}
}

@inproceedings{rasdaman1,
  title={The RasDaMan approach to multidimensional database management},
  author={Baumann, Peter and Furtado, Paula and Ritsch, Roland and Widmann, Norbert},
  booktitle={Proceedings of the 1997 ACM symposium on Applied computing},
  pages={166--173},
  year={1997}
}

@inproceedings{rasdaman2,
  title={The multidimensional database system RasDaMan},
  author={Baumann, Peter and Dehmel, Andreas and Furtado, Paula and Ritsch, Roland and Widmann, Norbert},
  booktitle={Proceedings of the 1998 ACM SIGMOD international conference on Management of data},
  pages={575--577},
  year={1998}
}

@article{deepesdl,
  title={{Deep Earth System Data Laboratory (DeepESDL)}},
  author={Anghelea, Anca and Dobrowolska, Ewelina and Brandt, Gunnar and Reinhardt, Martin and Mahecha, Miguel and Morbagal Harish, Tejas and Meissl, Stephan},
  journal={The International Archives of the Photogrammetry, Remote Sensing and Spatial Information Sciences},
  volume={48},
  pages={13--18},
  year={2024},
  publisher={Copernicus Publications G{\"o}ttingen, Germany}
}

@misc{EarthSystemDataLab_2025,
  title        = {Earth System Data Lab},
  author       = {{Deep Earth System Data Laboratory (DeepESDL) Team}},
  year         = {2025},
  howpublished = {\url{https://earthsystemdatalab.net/}},
  note         = {Accessed on 2025-12-30},
  url          = {https://earthsystemdatalab.net/},
  organization = {Earth System Data Lab / DeepESDL},
}

@article{big_data_challenges,
  title={{Big data in Earth science: Emerging practice and promise}},
  author={Vance, Tiffany C and Huang, Thomas and Butler, Kevin A},
  journal={Science},
  volume={383},
  number={6688},
  pages={eadh9607},
  year={2024},
  publisher={American Association for the Advancement of Science}
}

@article{big_data_frontier,
  title={{Big Earth data: A new frontier in Earth and information sciences}},
  author={Guo, Huadong},
  journal={Big Earth Data},
  volume={1},
  number={1-2},
  pages={4--20},
  year={2017},
  publisher={Taylor \& Francis}
}

@article{rolap_qube,
  title={ROLAP implementations of the data cube},
  author={Morfonios, Konstantinos and Konakas, Stratis and Ioannidis, Yannis and Kotsis, Nikolaos},
  journal={ACM Computing Surveys (CSUR)},
  volume={39},
  number={4},
  pages={12--es},
  year={2007},
  publisher={ACM New York, NY, USA}
}

@article{datacube1,
  title={Data cube: A relational aggregation operator generalizing group-by, cross-tab, and sub-totals},
  author={Gray, Jim and Chaudhuri, Surajit and Bosworth, Adam and Layman, Andrew and Reichart, Don and Venkatrao, Murali and Pellow, Frank and Pirahesh, Hamid},
  journal={Data mining and knowledge discovery},
  volume={1},
  number={1},
  pages={29--53},
  year={1997},
  publisher={Springer}
}

@article{datacube2,
  title={Implementing data cubes efficiently},
  author={Harinarayan, Venky and Rajaraman, Anand and Ullman, Jeffrey D},
  journal={Acm Sigmod Record},
  volume={25},
  number={2},
  pages={205--216},
  year={1996},
  publisher={ACM New York, NY, USA}
}

@inproceedings{eo_datacube1,
  title={Overview of the open data cube initiative},
  author={Killough, Brian},
  booktitle={IGARSS 2018-2018 IEEE international geoscience and remote sensing symposium},
  pages={8629--8632},
  year={2018},
  organization={IEEE}
}

@article{eo_datacube2,
  title={The Australian geoscience data cube—foundations and lessons learned},
  author={Lewis, Adam and Oliver, Simon and Lymburner, Leo and Evans, Ben and Wyborn, Lesley and Mueller, Norman and Raevksi, Gregory and Hooke, Jeremy and Woodcock, Rob and Sixsmith, Joshua and others},
  journal={Remote Sensing of Environment},
  volume={202},
  pages={276--292},
  year={2017},
  publisher={Elsevier}
}

@article{eo_datacube3,
  title={Building an earth observations data cube: lessons learned from the swiss data cube (sdc) on generating analysis ready data (ard)},
  author={Giuliani, Gregory and Chatenoux, Bruno and De Bono, Andrea and Rodila, Denisa and Richard, Jean-Philippe and Allenbach, Karin and Dao, Hy and Peduzzi, Pascal},
  journal={Big Earth Data},
  volume={1},
  number={1-2},
  pages={100--117},
  year={2017},
  publisher={Taylor \& Francis}
}

@article{eo_datacube4,
  title={Monitoring land degradation at national level using satellite Earth Observation time-series data to support SDG15--exploring the potential of data cube},
  author={Giuliani, Gregory and Chatenoux, Bruno and Benvenuti, Antonio and Lacroix, Pierre and Santoro, Mattia and Mazzetti, Paolo},
  journal={Big Earth Data},
  volume={4},
  number={1},
  pages={3--22},
  year={2020},
  publisher={Taylor \& Francis}
}

@article{medical_datacube,
  title={Privacy-preserving data cube for electronic medical records: An experimental evaluation},
  author={Kim, Soohyung and Lee, Hyukki and Chung, Yon Dohn},
  journal={International journal of medical informatics},
  volume={97},
  pages={33--42},
  year={2017},
  publisher={Elsevier}
}

@article{social_datacube,
  title={Development of multidimensional academic information networks with a novel data cube based modeling method},
  author={Kaya, Mehmet and Alhajj, Reda},
  journal={Information Sciences},
  volume={265},
  pages={211--224},
  year={2014},
  publisher={Elsevier}
}

@inproceedings{network_datacube,
  title={Graph cube: on warehousing and OLAP multidimensional networks},
  author={Zhao, Peixiang and Li, Xiaolei and Xin, Dong and Han, Jiawei},
  booktitle={Proceedings of the 2011 ACM SIGMOD International Conference on Management of data},
  pages={853--864},
  year={2011}
}
\end{document}